\documentclass[]{pnastwo}

\usepackage{epsf}
\usepackage{graphicx}
\usepackage{float}
\usepackage{afterpage}
\usepackage{relsize}
\usepackage{bm,bbm,amssymb,color, amsmath, color}
\usepackage{setspace}
\newcommand{\bra}[1]{\langle #1|}
\newcommand{\ket}[1]{|#1\rangle}
\newcommand{\inner}[2]{\langle #1|#2 \rangle }

\newcommand{\eq}[1]{\begin{align}
#1
\end{align}}

\begin{document}

\title{\bf\sf\LARGE Regular neutral networks outperform robust ones by reaching their top growth rate more quickly.}
\author{\sf J R Blundell\affil{1}{London Institute for Mathematical Sciences, London, UK}\affil{2}{Centre national de la recherche scientifique, Paris, France}\affil{3}{Cavendish Laboratory, Cambridge, UK},
\sf A Gallagher\affil{1},\sf\and \sf T M A Fink \affil{1}{}\affil{2}{}}
%

\maketitle

\begin{article}

\begin{abstract}
We study the relative importance of ``top-speed" (long-term growth rate) and ``acceleration" (how quickly the long-term growth rate can be reached) in the evolutionary race to increase population size.  We observe that fitness alone does not capture growth rate: robustness, a property of neutral network shape, 
combines with fitness to include the effect of deleterious mutations,
giving growth rate. 
Similarly, we show that growth rate alone does not capture population size:
regularity, a different property of neutral network shape, 
combines with growth rate to include the effect of higher depletion rates early on,
giving size. 
Whereas robustness is a function of the principal eigenvalue of the neutral network adjacency matrix,
regularity is a function of the principal eigenvector. 
We show that robustness is not correlated with regularity, and observe {\it in silico} the selection for regularity by evolving RNA ribozymes.
Despite having smaller growth rates, the more regular ribozymes have the biggest populations.
\end{abstract}


\section{Introduction}

\noindent
What wins a race, acceleration or top speed? 
In a long race, it's top speed; in a short race, acceleration. 
In the evolutionary race to increase population size, an organism's ``top speed'' (long-term growth rate) depends on its \emph{robustness}: 
a property of its underlying neutral network which determines the fraction of mutations that are selectively neutral \cite{Nimwegen_1999, Wagner_2008}.

In a changing environment, however, the race  is short. 
Then what matters is not ``top speed" but ``acceleration'' (how quickly a population achieves its long-term growth rate). We find that over shorter times, or in a changing environment, the most successful organisms are those which are able to reach their top growth rate quickly, even if they ultimately grow with a lower rate. 

{\it Robustness}. Ever since Kimura's initial work on neutral mutations in evolution \cite{Kimura_1964}, the role of robustness in determining population fate has been the subject of intense research \cite{Kimura_1991, Nimwegen_2}.  {\it In vitro} studies on RNA and protein evolution \cite{Wright, Bershtein_2006}, analyses of molecular codes \cite{Maeshiro_1998, Tlusty}, and mounting evidence from {\it in silico} RNA evolution \cite{Nimwegen_1999, Wagner_2008} have highlighted that robustness plays an important role in an organism's capacity to survive and, strikingly, to adapt \cite{Wagner_2008, Gunter_2, Bloom_2006, Draghi_2010, Plotkin_2}. 
Recent work suggests these observations also apply to non-biological models of self-assembly \cite{Johnston_2011, Ard_2} and programmable hardware \cite{Wagner_2010}. 

A detailed understanding of the relation between fitness and robustness in the long-time limit was put forward over a decade ago \cite{Nimwegen_1999}. This quantified how robustness affects an organisms long-term growth rate and led to the realisation that robustness could sometimes be more important than fitness itself, accounting for the qualitative distinction between ``survival of the fittest" and ``survival of the flattest"  \cite{Wilke_2001, Sole_2008}.

{\it Regularity}. 
We show that an organism's ``acceleration" depends on its \emph{regularity}: a new property of its neutral network which determines how quickly a population reaches its steady-state growth rate. Starting from a non-equilibrium distribution of population, regularity adds up the population loss due to deleterious mutations in excess of the population's long-term rate of loss.
For some neutral networks, this excess loss in the run-up to steady state can decimate the population size.
We demonstrate this effect for small neutral networks and, by evolving the Hammerhead ribozyme \textit{in silico}, for large ones.

\begin{figure}[b!]
\begin{center}
\includegraphics[scale=1.11]{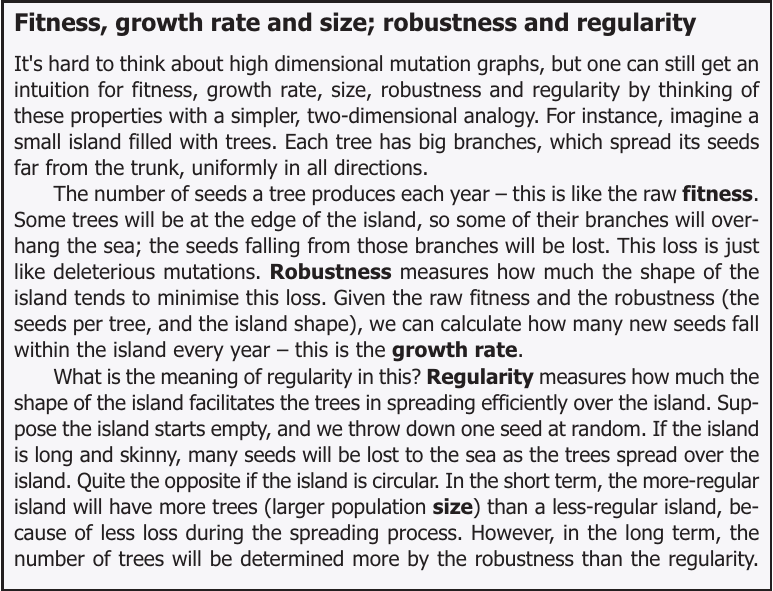}\end{center}
\end{figure}

In this paper we do the following six things, corresponding to the subsequent six sections:
\begin{enumerate}

\item We re-derive the infinite-time population growth rate $h$ in terms of the fitness $f$ and robustness $r$: $h = f (1 \! - \! d \mu (1\!-\!r))$.
We show that $r$ is set by the principal eigenvalue of its neutral network. Both were first done in \cite{Nimwegen_1999}.

\item We derive the finite-time population size $N$ in terms of growth rate $h$ and regularity $q$: $N(t) =  N(0) q h^t$. We show $q$ is set by the principal eigenvector of its neutral network.

\item We calculate the critical mutation rate $\mu_c$ which separates the regime where higher fitness wins from the one where higher robustness wins: $\mu_c = {(f_1/ f_2 - 1)}/({d (r_2 - r_1)})$. We verify this crossover for simple neutral networks.

\item We calculate the critical time $t_c$ which separates the regime where higher regularity wins from the one where higher growth rate wins:
$t_c = \ln(q_2/q_1)/\ln(h_1/h_2)$. 
We verify this crossover for simple and complex neutral networks.

\item We provide numerical and analytic evidence that regularity $q$ and robustness $r$ are uncorrelated. We construct neutral networks with low $q$ and high $r$, and high $q$ and low $r$.

\item We confirm that regularity is subject to selection on short time scales by simulating the evolution of the Hammerhead RNA in competition with a mutant phenotype.

\end{enumerate}

\newpage
\section{Robustness}
\noindent {\it Mutation graphs.} 
We study a generalized genome (the set of all genotypes) of length $d$ and alphabet size $a$. We associate with the genome a mutation graph, where each of the $a^d$ genotypes corresponds to a vertex and two vertices are connected by an edge if the genotypes differ by a single mutation. Each vertex is connected to $d(a -1)$ edges. In much of this paper we take $a = 2$ for simplicity but our results extend to alphabets of arbitrary size. We colour the vertices according to phenotype, where all
genotypes in a phenotype have the same colour and belong to the same neutral network (Fig.\ 1). 

{\it Mutational flux.} Mutation induces a population flux across neighbouring genotypes. If the mutation rate per nucleotide is $\mu$, and the genome length is $d$, the mutational flux is $1-(1-\mu)^d \simeq \mu d$ for $\mu d \ll 1$. It is the fraction of a population that mutates per generation. 
Some of this mutational flux will also cross phenotypic boundaries when neighbouring genotypes lie in two different phenotypes.
The rest of the flux is neutral.

{\it Genotype robustness.}
The genotype robustness $r_i$ \cite{Wagner_2008} is the probability of a mutation being neutral, that is, the fraction of edges leaving genotype $g_i$ which do not lead to a different phenotype.
It is the number of neutral edges $k_i$ divided by the total number of edges $d$,
\begin{equation}
 r_i = k_i / d.
\end{equation}

{\it Phenotype robustness.}
Consider a neutral network $P$ (possibly composed of disjoint clusters), of size $s$ (there are $s$ genotypes in $P$) and whose genotypes have fitness $f$. Let $N(t)$ be the total population on $P$. It is distributed over genotypes $g_i$ in $P$ according to $n_i(t)$. The normalized population is distributed according to $w_i(t) = n_i(t)/N(t)$ and is the fraction of the population on genotype $g_i$. 
In the large-time limit, the population is distributed according to a unique distribution ${w}_i(\infty)$, which we abbreviate $\tilde{w}_i$.
We define phenotype robustness $r$ to be the large-time popuation-weighted average of the genotype robustness $r_i$:
\eq{r = \sum_{i=1}^{s} \tilde{w}_i r_i. \label{def_rob} }
It is the fraction of the population flux that is neutral.
Note that $r \geq \langle r_i \rangle$, where $\langle r_i \rangle = {1\over s} \sum_i r_i$
is the mean value of $r_i$. This is because the population tends to concentrate in the network interior, away from surface genotypes with low $r_i$. 

{\it Fitness.} The fitness $f$ is the raw reproductive rate of the phenotype.
After $t$ generations, the total population $N(t)$ of a neutral network will have changed by a factor of $f^t$, in the absence of mutations.


{\it Structure factor.} 
At large time $t$, at every generation, a fraction $\mu d (1-r)$ of the population mutates off the neutral network. If we assume that  neighbouring phenotypes have negligible fitness, then at each generation the population $N(t)$ is multiplied by $\gamma$, in addition to its inherent fitness, where
\begin{equation}
\gamma = 1- d \mu (1 - r),
\end{equation}
which we call the {\it structure factor}. It depends on the shape of the neutral network. While $f$ may cause the population to increase or decrease,  $\gamma$ can only decrease it or leave it as is.

{\it Growth rate.}
The overall factor by which a mutating population changes over the span of a generation is the product of its fitness $f$ and its structure factor $\gamma$,
\eq{
h = f \gamma = f(1-d \mu(1-r))
\label{free_fitness}}
where we call $h$ the {\it growth rate}.
The growth rate is that fitness that can be usefully employed to increase the population and not spent replenishing population lost to deleterious mutations incurred at the boundary; $h \leq f$. For a single neutral network taken in isolation, it is $h$ rather than $f$ which will determine whether and to what extent the population will expand or diminish. 

\begin{figure}[b!]
\begin{center}
\includegraphics[scale=0.35]{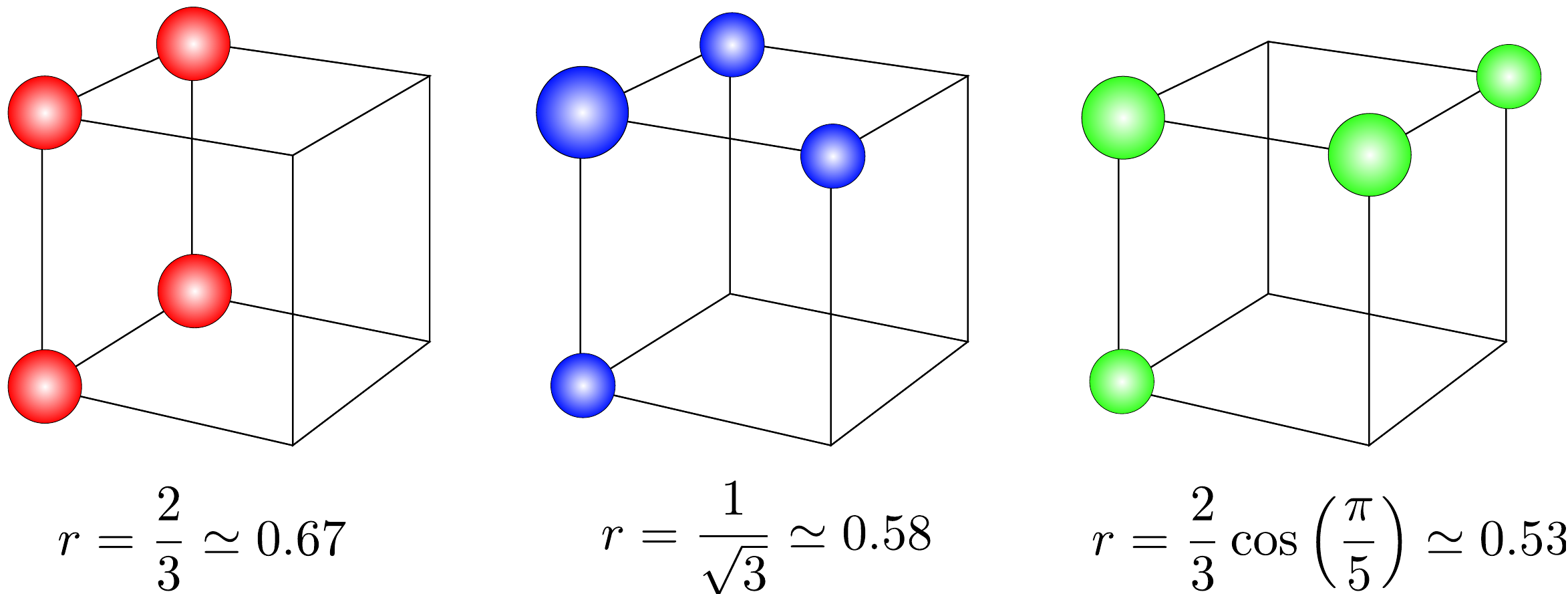}
\caption{Robustness. The three connected neutral networks of size $s$=4 on the 3-cube ($d$=3, $a$=2), in order of robustness. For a given neutral network, the area of a vertex is proportional to the fraction of population on it, in the large-time limit.}  \label{figure1}
\end{center}
\end{figure}


{\it Dirac notation}. 
For the rest of this section and the next we use Dirac, or bra-ket, notation, standard in quantum mechanics.
A vector is denoted by $\ket{x}$ and its transpose by $\bra{x}$. 
The inner and outer products of $\bra{x}$ and $\ket{y}$ are denoted by $\inner{x}{y}$ and $\ket{y}\bra{x}$. 
Let the genotype selection vector $\ket{g_i} = [0,\ldots,0,i,0,\ldots,0]^T$; it selects that component of a vector which projects onto genotype $i$.

{\it Mutation matrix}. 
The action of mutation on the population distribution over a single generation can be expressed by the mutation matrix $M$: 
\eq{
M_{ij}(\mu) = (1-d \mu) \delta_{ij} + \mu A_{ij}.
\label{K}
}
Here $A_{ij}$ is the adjacency matrix of the neutral network: $A_{ij} = 1$ if nodes $i$ and $j$ share an edge and $A_{ij} = 0$ otherwise. 
The first term is the probability that no mutation occurs and the second the probability of mutating from $i$ to $j$. 
Being symmetric, $M$ can be diagonalised by its eigenvectors $\ket{e_i}$:
\eq{
M = (1-\mu d)\sum_i \ket{e_i} \bra{e_i} + \mu \sum_i \ket{e_i} \lambda_i\bra{e_i},
\label{mutation_matrix}
}
where the  $\ket{e_i}$ satisfy $M \ket{e_i} = \lambda_i \ket{e_i}$ and $\inner{e_i}{e_j}=\delta_{ij}$. 

\textit{Population vector}. The population vector $\ket{x_t}$ gives the size and distribution of the population. Its component $\inner{g_i}{x_t}$ is the population $n_i(t)$ on $g_i$, where $\inner{g_i}{g_j}=\delta_{ij}$. 
The population vector $\ket{x_t}$ is obtained by transforming an initial vector $\ket{x_0}$ by $M^t$ and multiplying it by $f^t$:
\eq{
\ket{x_t} = M^t f^t \ket{x_0} &=  \sum_i \inner{e_i}{x_0} \big(1-\mu d \left(1-\lambda_i/d\right)\big)^t f^t \ket{e_i}  \label{population_distribution}
}
Since $|\lambda_i| \leq d$,  all terms $i>1$ decay exponentially with respect to the first for $\mu>0$. 
In the large time limit the sum is dominated by the first term, whose eigenvalue $\lambda_1$ is largest:
\eq{
\ket{x_t} =  \inner{e_1}{x_0} f^t (1- d \mu (1 - r))^t \ket{e_1} =  \inner{e_1}{x_0}h^t \ket{e_1}
\label{x_t_large_time_limit}
}
where we have defined the phenotype robustness to be
\eq{
r = \lambda_1/d.
\label{r_eigen_def}
}
The quantity $r$ measures how well the shape of the neutral network $P$ can reduce the rate of deleterious mutation acting on the population as a whole. Since the eigenvalues depend only on the adjacency matrix $A$, the steady state population distribution depends only on the shape of $P$ and on neither the mutation rate $\mu$ nor the fitness $f$ \cite{Nimwegen_1999}.

\newpage
\section{Regularity}
\noindent
Here we introduce a new property of neutral networks, {\it regularity}, which measures the reduction in population size whilst a population evolves out of equilibrium towards steady-state starting from a uniform distribution. We show that the regularity can have a dramatic effect on population size in the short term, long before steady state is reached.

{\it Population size.}
In the previous section we calculated the population {\it growth rate} at infinite time (steady state) by computing $|f^t M^t\ket{x_0} = \ket{x_t}$. 
Here we explicitly calculate the population {\it size} at finite time by computing 
\eq{
N(t) = \sum_i  \bra{g_i} f^t  M^t \ket{x_0} = \sum_i  \inner{g_i}{x_t}.
}
Substituting $\ket{x_t}$ from (\ref{x_t_large_time_limit}) into the above yields
\eq{
N(t)= h^t \inner{e_1}{x_0} \sum_i \inner{g_i}{e_1};
\label{population_size}
}
the population size depends on the initial distribution $\ket{x_0}$. 

{\it Simulation}. 
In Fig.\ 3 we plot the population size as a function of time for two neutral networks, $P_1$ (top) and $P_2$ (bottom). 
In both, the initial condition $\ket{x_0}$ is a single adaptive mutant: the population is confined to a single genotype (grey lines). 
In $P_1$, the long-term population distribution is very non-uniform. Adaptive mutants starting from barren genotypes fare poorly, with long-term population sizes $10^{-4}$ the size of the best performing genotypes. 
In $P_2$, the long-term population distribution is comparably uniform, and the average adaptive mutant fares much better. 
We explain this phenomenon in terms of the regularity below.

In the run-up to steady state, a uniform population distribution becomes more and more robust. 
Accordingly, more of the population is depleted at early generations than at later ones.
The difference between the depletion rates at finite time and at infinite time $d \mu (1-r)$ is the {\it excess depletion}.
The regularity $q$ is the geometric reduction in population size due to the cumulative excess depletion,  starting from a uniform population distribution.
In other words, $q$ is the ratio of the long-term population size that develops when $\ket{x_0}$ is a uniform distribution $\ket{u}$ and when $\ket{x_0}$ is the principal eigenvector $\ket{e_1}$. 

\begin{figure}[b!]
\begin{center}
\includegraphics[scale=0.33]{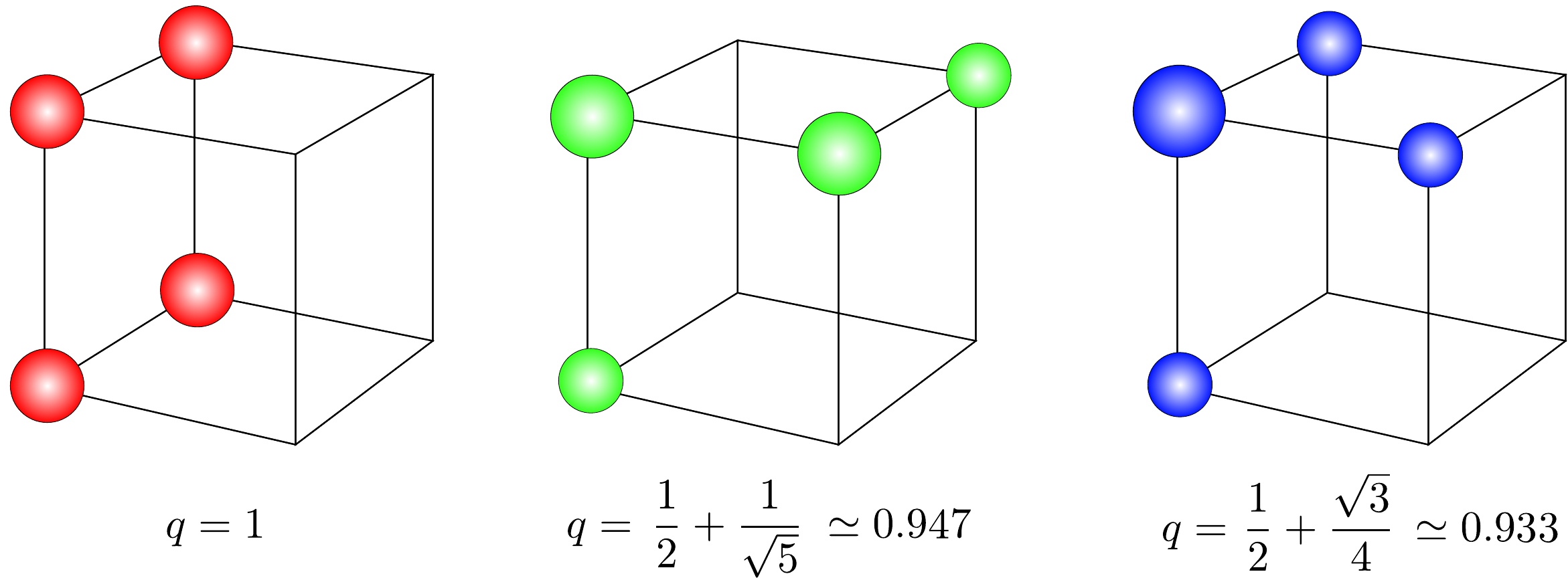}
\caption{Regularity. The three connected neutral networks of size $s$=4 on the 3-cube ($d$=3, $a$=2), in order of regularity. For a given neutral network, the area of a vertex is proportional to the fraction of population on it, in the large time limit.}
\label{lowerbound}
\end{center}
\end{figure} 

The uniform distribution is $\ket{u} = N_0/s\sum_i \ket{g_i}$. Replacing $\ket{x_0}$ with $\ket{u}$ in (\ref{population_size}) yields
\eq{
N(t) = N_0 h^t  \frac{1}{s}\left(\sum_i \inner{g_i}{e_1}\right)^2. 
\label{uniform_initial}
}
The steady-state distribution is $N_0 \ket{e_1} / \sum_i \inner{g_i}{e_1}$. 
Substituting it into (\ref{population_size}) yields
\eq{
N(t) = N_0 h^t.
\label{steady_state_initial}
}
Taking the ratio of (\ref{uniform_initial}) and (\ref{steady_state_initial}), we obtain for the regularity
\eq{
q=\frac{1}{s}\left(\sum_i \inner{g_1}{e_1}\right)^2.
\label{reg_defined}
}

For an alternative but equivalent definition of regularity, imagine instead that the neutral network is discovered by a single adaptive mutant, with the initial condition $\ket{x_0} =  \ket{g_i}$. 
Assuming all genotypes in the phenotype are equally likely to be the ``port of entry," and recalling that
the uniform distribution is the sum over all single adaptive mutants, 
we see that the fate of $\ket{x_0} = \ket{u}$ is the mean of the fates of $\ket{x_0} = \ket{g_i}$, averaged over all $g_i$ in $P$.

Finally, we can define the regularity $q$ in terms of the population distribution $\tilde{w}_i$. Recall that
\eq{
\tilde{w}_i = \frac{\inner{g_i}{e_1}}{\sum_i \inner{g_i}{e_1}}.
}
Squaring both sides and summing over $i$ we find $\sum_i \tilde{w}_i^2 = 1/qs$,
having used the identity $\inner{e_1}{e_1} = \sum_i \inner{g_i}{e_1}^2 = 1$. Then 
\eq{
q = \frac{1}{s^2 \langle \tilde{w}_i^2 \rangle };
}
the regularity is proportional to the mean square of the population distribution at large time.
Since the mean square of a normalised distribution is maximised when all the weight is on a single point, and minimised when the distribution is uniform, we observe that $q$ satisfies the bounds
\eq{
1/s < q \leq 1.
}
The left relation is an inequality because it is not possible for a connected neutral network of size 2 or more to have an eigenvector confined to a single genotype. Note that for large neutral networks the effect of regularity can be dramatic since $1/s$ can be very small.

 
\begin{figure}[b!]
\begin{center}
\includegraphics[scale=0.48]{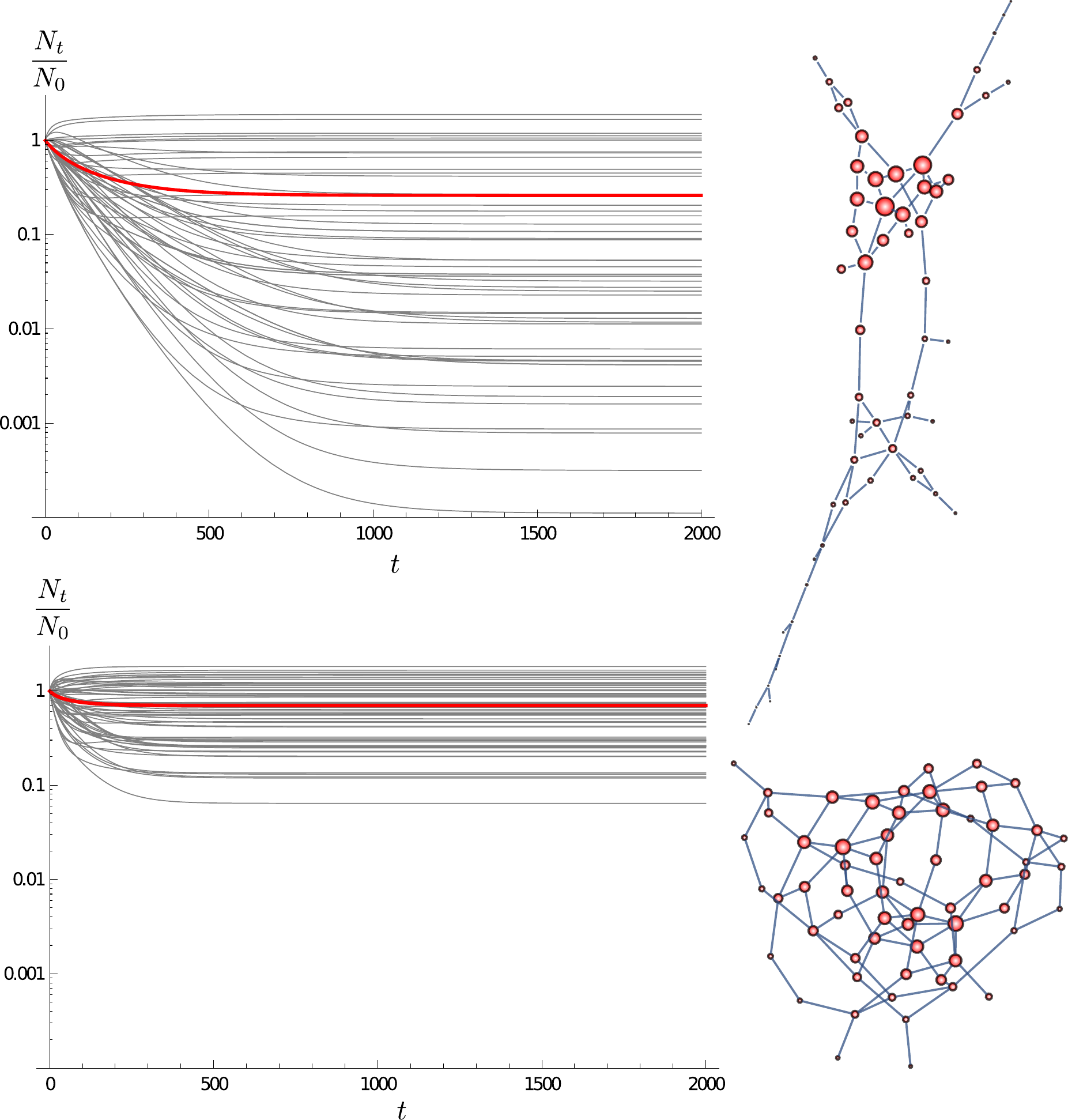}
\caption{Regularity. Two neutral networks drawn from an 8-cube ($d=8, a=2$) and with growth rate $h=1$. 
The top network has low regularity ($q \simeq 0.25$); the bottom has high ($q \simeq 0.70$). 
Grey lines show the evolution of population size, starting from unit size at a single genotype. 
The population growth rate at steady state is unity, but the population size at steady state varies radically. 
The red line is the mean population size; this is equivalent to the population size starting from a uniform initial distribution. 
At large times, the red line is the regularity. 
For the irregular network (top), the population can diminish by the factor $10^{-4}$.}
\label{regularities}
\end{center}
\end{figure}

\newpage
\section{Crossover from fitness to robustness}
\noindent
In this section we quantify the transition from fitness dominance to robustness dominance as a function of mutation rate $\mu$.
After a long time when the population is in steady state, it reproduces according to the growth rate
\eq{
h= f(1-\mu d (1-r)), 
\label{free_fitness2}
}
where the robustness $r = \lambda_1/d$ and $\lambda_1$ is the principal eigenvalue of the network adjacency matrix.

{\it Crossover.} Equation (\ref{free_fitness2}) predicts the onset of so-called ``survival of the flattest" \cite{Wilke_2001, Sole_2008} for sufficiently large mutation rates $\mu$.
For example, two neutral networks $P_1$ and $P_2$ with $f_1 > f_2$ and $r_2>r_1$ can show a crossover whereby the more fit network wins at low mutation rate $\mu$, but the more robust network wins for large $\mu$. The crossover occurs when $h_1 = h_2$, from which we see that the exchange rate between fitness and robustness is
\eq{
f_1/f_2 -1 = \mu d (r_2 - r_1).
\label{exchange_rate}
}
Fixing all but $\mu$ and rearranging, the critical mutation rate $\mu_c$ at which we observe a cross-over from ``survival of the fittest" to ``survival of the flattest" is
\eq{
\mu_c = \frac{f_1/ f_2 - 1}{d (r_2 - r_1)}.
}
For $\mu < \mu_c$ the more fit network wins despite being less robust;
for $\mu > \mu_c$ the more robust wins despite being less fit.

{\it Simulation.}
We illustrate this effect in Fig.\ 4, where we simulated five different neutral networks of size 6 drawn from the 5-cube ($d=5, a=2$). 
Because the five network fitnesses are not in the same rank order as their robustnesses, they show a number of crossovers as mutation rate $\mu$ increases. 

{\it Comment.}
Note that $\mu d$, the mutation rate per genotype, plays the role of an effective temperature, governing the transition from rewarding fitness to rewarding flatness. 
This is analogous to the $-TS$ term in classical thermodynamics, there $T$ is temperature and $S$ is entropy. 
At $d \mu = 0$, robustness has no impact on the dynamics, and $h = f$; fitness alone is rewarded. 

\begin{figure}[b!]
\begin{center}
\includegraphics[scale=0.47]{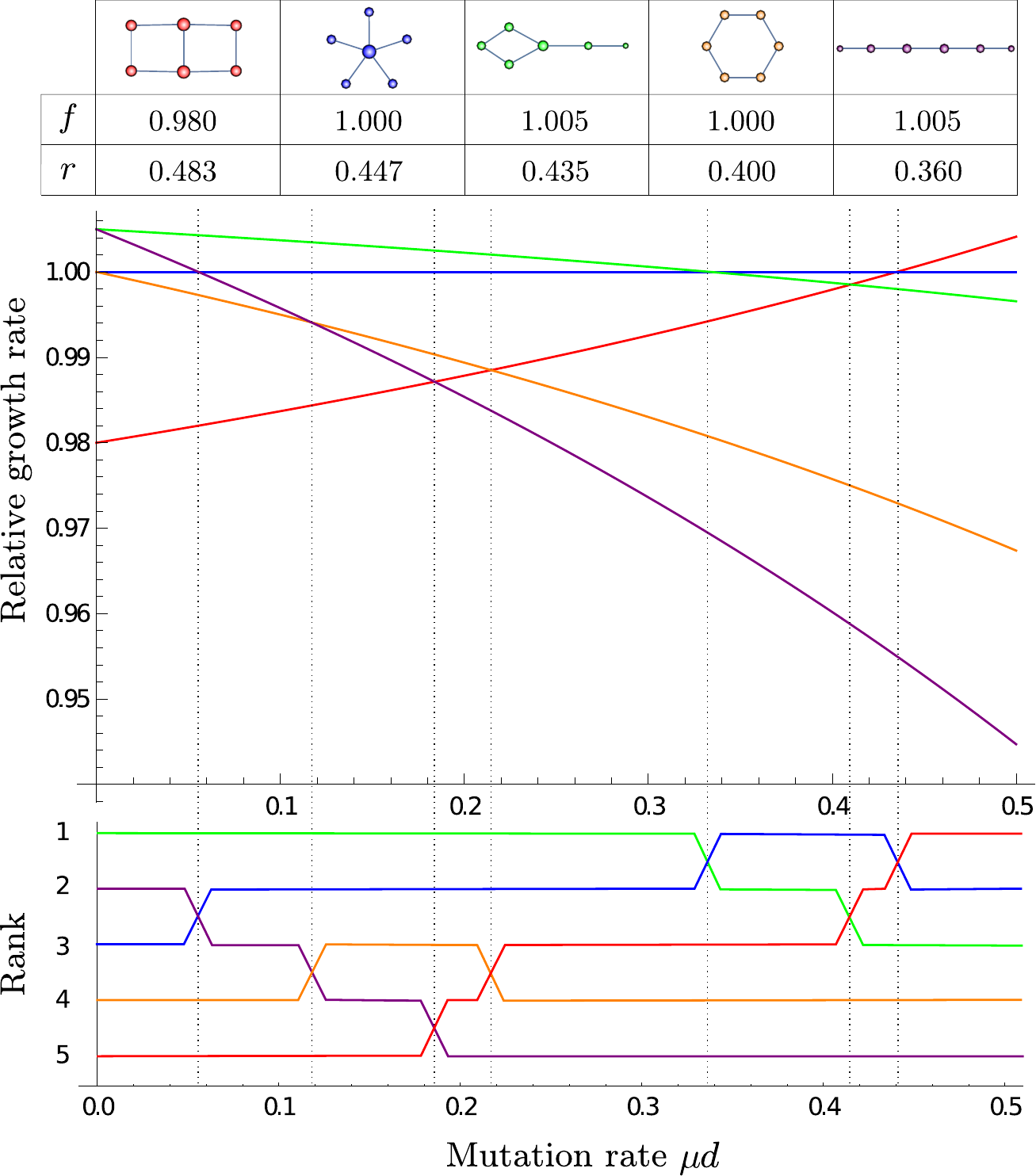}
\caption{Crossover from fitness to robustness. The growth rate $h = f (1- \mu d (1-r))$ of neutral networks whose fitness rankings are in a different order to their robustness rankings can show cross over from ``survival of fittest", to ``survival of flattest". In this case at high mutation rates the positive effect of increased robustness $r$ can outweigh the negative effect of a reduced fitness $f$.}
\label{fittest_flattest}
\end{center}
\end{figure}


\section{Crossover from growth rate to regularity}
In this section we quantify the transition from regularity dominance to growth rate dominance as a function of time $t$.
The population at steady state, starting from a uniform distribution (equally, averaged over all single adaptive mutants) is 
\eq{
N(t) = N_0 h^t q 
\label{shezzle}
}
where the regularity $q=\frac{1}{s}\left(\sum_i \inner{g_1}{e_1}\right)^2$ and $e_1$ is the principal eigenvector of the network adjacency matrix.

{\it Crossover.} Equation (\ref{shezzle}) predicts the transition to ``survival of the most regular" at sufficiently short times $t$. Again, two neutral networks $P_1$ and $P_2$ with $h_1 > h_2$ and $q_2>q_1$ can show a crossover whereby the more regular network wins at small (finite) time $t$, but the higher growth rate wins for large $t$. 
The crossover occurs when $N_1 = N_2$, and therefore the exchange rate between growth rate and regularity is
\eq{
q_2/q_1 = (h_1/h_2)^t.
\label{exchange_rate_h_q}
}
Fixing everything but $t$ and rearranging, the critical time $t_c$ at which we observe a cross-over from ``dominance of most regular" to ``dominance of highest growth rate" is
\eq{t_c = \frac{\ln(q_2/q_1)}{\ln(h_1/h_2)}}
For $t<t_c$ the more regular network has a larger population, while for $t>t_c$ the highest growth rate network dominates. 

{\it Simulation.}
We again illustrate this effect in Fig.\ 5, where we simulated five different neutral networks of size 6 drawn from the 5-cube ($d=5, a=2$). 
Because the five network growth rates are not in the same rank order as their regularities, they show a number of crossovers as time $t$ increases. 

{\it Comment.}
Where fitness and robustness combined to give a full picture of growth rate, now growth rate and regularity combine to give a full picture of population size.
These effects refer to the \textit{mean} size of an evolving population averaged over all single adaptive mutants. As shown in Fig.\ 3, the variance around this mean can be large and the population of the less regular network can be reduced by up to
\eq{N(t) = N_0 h^t \min \left( \inner{g_i}{e_1}^2\right)} 
which can be very small indeed.
\noindent


 \begin{figure}[b!]
\begin{center}
\includegraphics[scale=0.45]{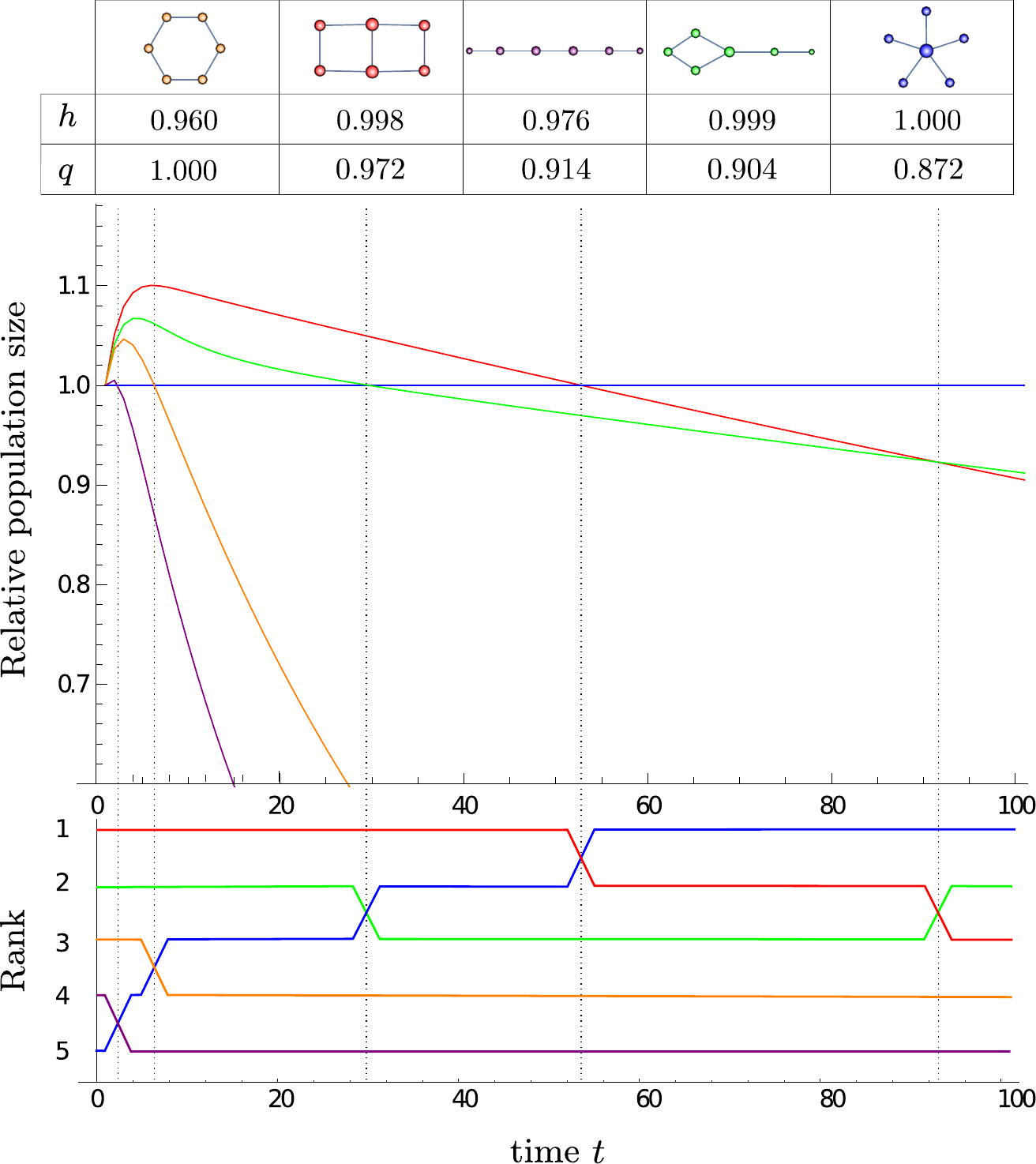}
\caption{Crossover from growth rate to regularity. Population sizes (relative to the blue one) as a function of time for neutral networks whose growth rate ranking are different from their regularity rankings can show crossovers. Here growth rates correspond to $\mu = 0.4$ in Fig.\ 4. Over short time more regular networks (orange, red) can win out over irregular ones (blue, green) despite being at a selective disadvantage. }
\label{robustness_regularity}
\end{center}
\end{figure}

\newpage
\section{Regularity is not robustness}

\noindent
Our simulation in Fig.\ 3 and others like it suggest that regular neutral networks, like robust ones, are highly connected.
Therefore it might seem that regularity is largely determined by robustness.
We show in this section, however, that for all but the smallest values of $d$, regularity $q$ is not correlated with robustness $r$, 
and for most values of $r$, there exist neutral networks with a broad range of $q$.

{\it Simulation.} 
We studied the relation between robustness $r$ and regularity $q$ by enumerating all neutral networks for $d=4$, 
and sampling $10,000$ neutral networks for $d=5$ and $d=6$; in all cases $a=2$.  
(The number of neutral networks grows rapidly with $d$, and calculating the principal eigenvectors for each gets more expensive.)  
The networks were constructed as follows: 
(i) with a uniform probability, each of the genotypes was selected to be part of the neutral network;
(ii) the uniform probability was slightly increased, and we repeated.

The results of the simulation are shown in Fig.\ 6, in a scatter plot of $r$-$q$ space. 
For $d=4$ (red points), $r$ varies from 0 to 1, but $q$ only varies from 0.77 to 1; there are no very irregular subgraphs of a 4-cube.
For $d=5$ (blue points), $q$ dips further down to 0.56, while for $d=6$ (grey points), $q$ dips to 0.42
 and more of the $r$-$q$ plane is filled. (In actuality, more of the plane is filled than shown due to under-sampling at the frontiers.)
In each case, as $d$ continues to increase, more and more of the unit plane is accessible, and for most $r$ there is a wide range of $q$.

{\it Tadpoles.} We can show for large $d$ that one can find neutral networks that span almost all of $r$-$q$ space.  A tadpole network is a $k$-dimensional hypercube with a path (a graph with no branches) appended to one corner.
For any tadpole, the principal eigenvalue $\lambda_1 > k$, since $\lambda_1=k$ for a $k$-cube and for any subgraph $U$ of $V$, 
$\lambda_1(U) < \lambda_1(V)$.
Then by (\ref{r_eigen_def}) the robustness $r$ of a tadpole satisfies $r > k/d$.
Whereas $r$ is dominated by the head of a tadpole, $q$ is dominated by the tail, since the population decays exponentially into the tail. This means that as one increases tail size for a given head size the regularity $q \sim 1/s$. Making the head of the tadpole larger increases both $r$ and $q$, but increasing the tail only reduces $q$, leaving $r$ almost unchanged. By choosing a head size $k$ such that $r = k/d$ and $q=1$, we can then reduce $q$  by increasing the size of the tail while having almost no effect on $r$, thereby forming almost any two values of $r$-$q$ required.


\section{Regularity is subject to selection}
\noindent 
In this section, by simulating the evolution of two RNA ribozymes, we provide further evidence that evolution can select for regularity at short time scales.

\begin{figure}[b!]
\begin{center}
\includegraphics[scale=0.85]{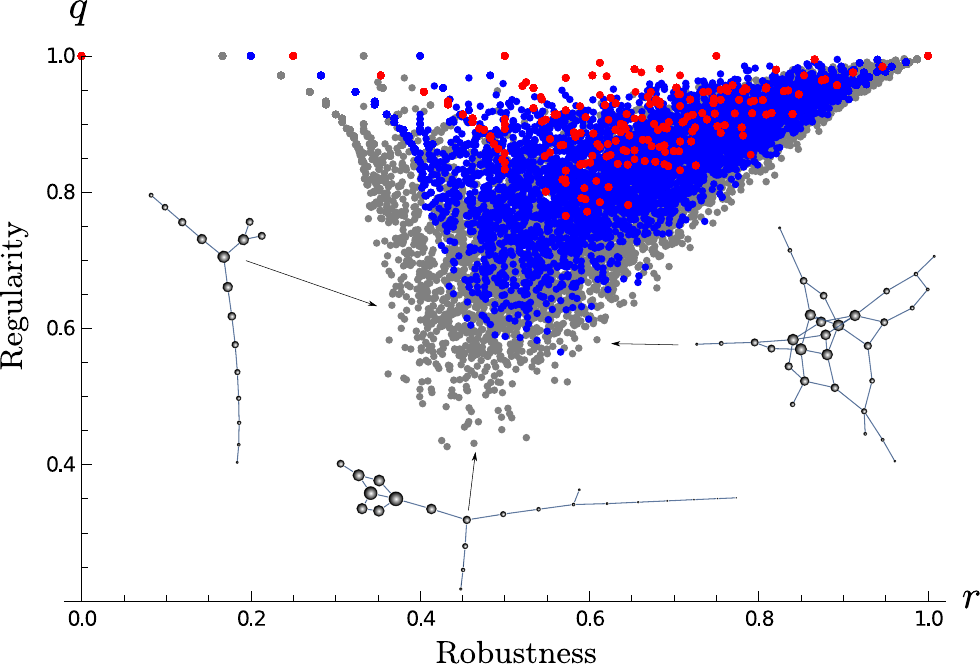}
\caption{Regularity is not robustness. A scatter-plot of robustness vs regularity for
all possible neutral networks for $d=4$ (red points), and
10,000 randomly sampled neutral networks for $d=5$ (blue points) and $d=6$ (grey points).
As $d$ increases, more of the $r$-$q$ unit square becomes accessible.
For three specific grey data points, the corresponding neutral networks are illustrated.}
\label{rob_not_reg}
\end{center}
\end{figure}

{\it Two phenotypes.}
We consider two RNA secondary structures in Fig.\  7. 
The first is the Hammerhead ribozyme (Ham);
the second is a mutant phenotype (Mut) with a considerably different secondary structure. Both are formed from an RNA of length $d=45$.
We assign both phenotypes the same fitness $f$. 
In order to visualise the neutral networks of Ham and Mut, we constrain mutation so that it can only occur at $6$ randomly chosen ``hotspots" on the sequence (black squares). 
Each network is a slice of its full neutral network along $6$ of the possible $45$ dimensions. 
Vertex area is proportional to the steady-state population distribution, $\tilde{w}_i$.
Of the $4^6$ allowed sequences (genotypes), $38$ fold to Ham (Fig.\ 7, left)
and $141$ fold to Mut (Fig.\ 7, right); the rest fold to various different secondary structures, not shown. 
All folding was performed using the ViennaRNA package \cite{ViennaRNA_1}.
Ham has lower robustness $r$ than Mut (0.33 vs 0.46) and therefore lower growth rate $h = f(1-d \mu(1-r))$, but a higher regularity $q$ (0.97 vs 0.67).

{\it Simulation.} 
We simulated the evolution of a population of $N = 1,000$ sequences, equally split between Ham and Mut ($N_H = N_M = 500$). 
For each run, the Ham population $N_H$ all began on the same genotype, randomly chosen from its neutral network; 
the same applies to Mut and its network.
At every generation, for each sequence a point mutation occurred somewhere along the chain with probability $d (a-1) \mu=18 \mu=0.1$, 
and the sequence survived if its phenotype was preserved, but died if it was not.
Then the total population was renormalized by randomly selecting $1000$ sequences with replacement.    

{\it At steady state.}
After a long time when steady state is reached, the story is simple:  
$1-\mu d(1-0.33)$ of the Ham population advanced to the next generation, whereas 
$1-\mu d(1- 0.46)$ of the Mut population did so.
The selective advantage conferred by the higher robustness meant that Mut always won and Ham always lost over 100 runs (Fig.\ 8, left).

{\it Before steady state.}
At shorter time scales, the story is more subtle. 
Although Mut is more robust, it is less regular, and $10 \%$ of the time the Mut population suffered an early invasion from Ham (Fig.\ 8, right).
Ham ultimately reproduces with a lower rate than Mut, but {\it on average it gets to its top rate more quickly}.
In the Mut network (Fig.\ 7, right), there are $14$ genotypes located on the left peninsula of the network whose components 
$\tilde{w}_i < 10^{-3}$. 
If the Mut population begins on one of these genotypes, it is likely to suffer an early invasion by Ham. 
For our finite population, this invasion is sometimes large enough for Mut to go extinct due to drift. 

\begin{figure}[b!]
\begin{center}
\includegraphics[scale=0.3]{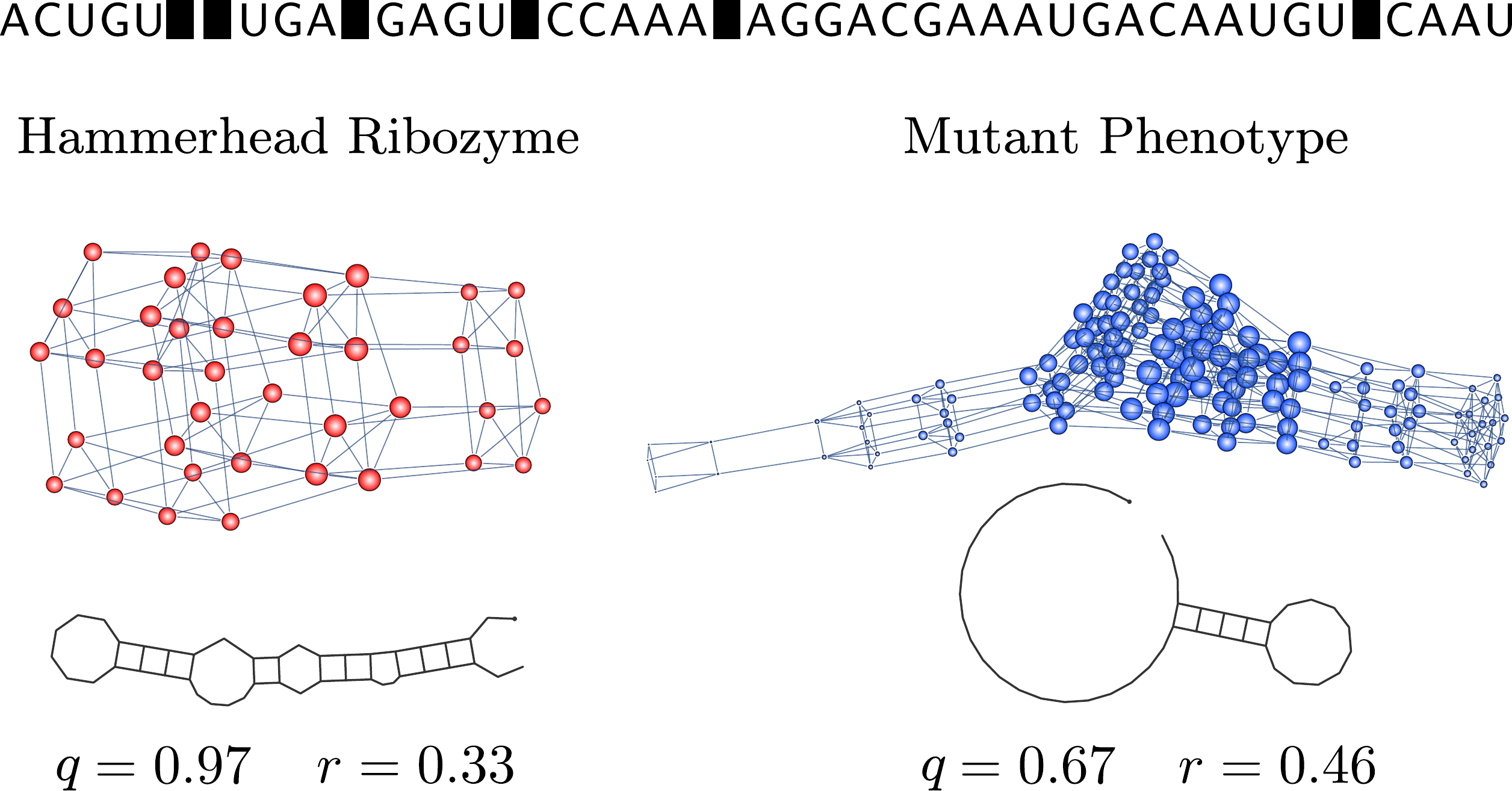}
\caption{Regularity is subject to selection.
(Top) RNA sequence showing 6 hotspots (black squares) where mutation is allowed. 
(Left) The neutral network formed by the 38 of the $4^6$ allowed sequences that fold to the Hammerhead ribozyme (Ham) secondary structure.
(Right) The same but for the 141 sequences that fold to a mutant (Mut) secondary structure. 
Vertex area is proportional to the long-time population distribution, $\tilde{w}_i$.
Mut is more robust than Ham, but less regular.
}
\label{RNA}
\end{center}
\end{figure}

\newpage

\begin{figure}[h!]
\begin{center}
\includegraphics[scale=0.33]{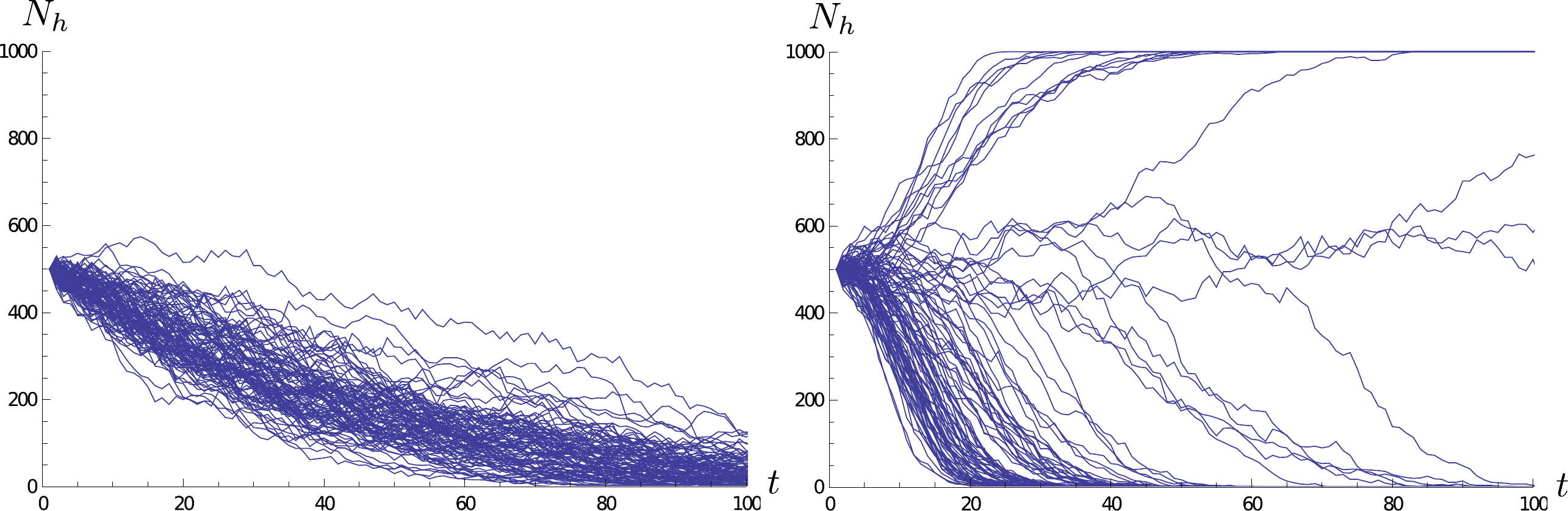}
\caption{
Regularity is subject to selection. Evolution of the of Hammerhead ribozyme secondary structure (Ham), in competition with a mutant structure (Mut). 
The Ham and Mut populations are normalised such that $N_H + N_M = 1,000$ and $\mu = 0.0056$.
(Left) At steady state, the more robust Mut wins every time out of 100 runs. 
(Right) At shorter times, the more regular Ham wins 10 times out of 100 runs.
Mut ultimately reproduces with a higher rate than Ham, but on average Ham gets to its top rate more quickly.
}
\label{RNA}
\end{center}
\end{figure}

\section{Conclusion}
 
The six main results of this paper are listed at the end of the Introduction.
Here we present some unifying observations, suggestions for experiment, and generalisations to other fields.


\begin{figure}[b!]
\begin{center}
\includegraphics[scale=0.32]{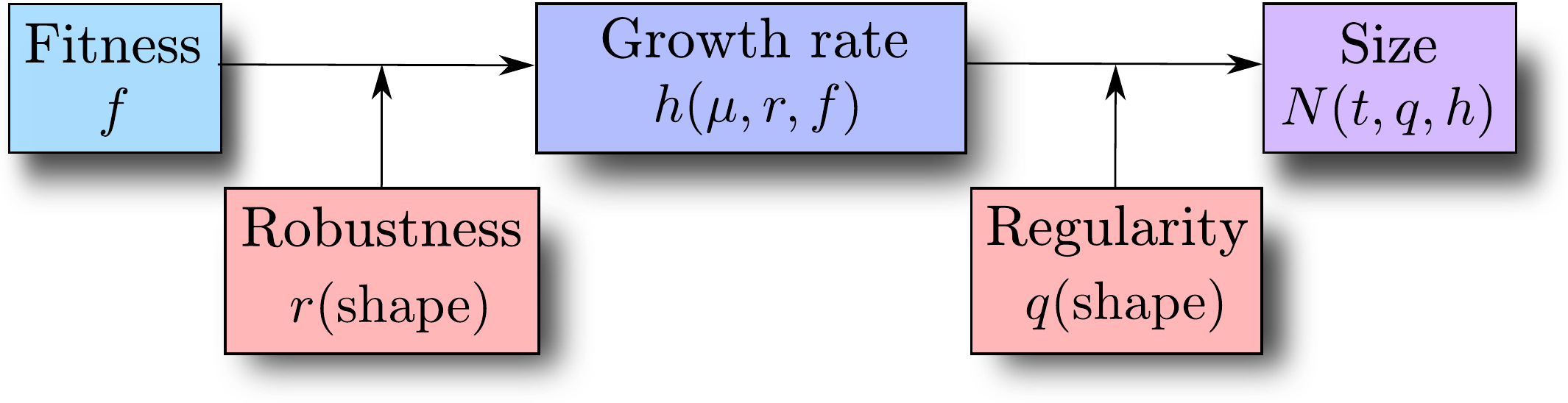}
\caption{Conclusion. 
The hierarchical relationships between fitness, growth rate and size.
As well as fitness and robustness, growth rate depends on the mutation rate.
As well as growth rate and regularity, size depends on time.
}
\label{quantities}
\end{center}
\end{figure}

{\it Fitness, growth rate and size.} 
The progression from fitness to growth rate to population size reflects both their chronology of discovery and their hierarchical relationship; 
this is illustrated in Fig.\ 9.
Fitness $f$ alone does not capture population growth rate. 
Robustness $r$, a property of neutral network shape, 
combines with fitness to include the effect of deleterious mutations,
giving the growth rate $h$. 
Similarly, growth rate $h$ alone does not capture population size. 
Regularity $q$, a different property of neutral network shape, 
combines with growth rate to include the effect of higher depletion rates early on,
giving the size $N$.
Both growth rate and size exhibit crossovers: 
from $f$ to $r$ as a function of mutation rate $\mu$, and
from $q$ to $h$ as a function of time $t$, respectively. 



{\it ``Smooth shapes are very rare in the wild,"} wrote Mandelbrot \cite{Mandelbrot}, ``but extremely important in the ivory tower and the factory."
The robustness $r$ and regularity $q$ characterise the shape of a graph, one via the graph's principal eigenvalue, the other its principal eigenvector.
But our intuition is valid for smooth, often Euclidean, shapes, not the jagged, high-dimensional shapes of neutral networks, which are themselves subgraphs of Hamming graphs. 
We have presented evidence that $r$ and $q$ are largely uncorrelated, but their behaviour is far from intuitive, and their precise relationship remains an open question.
We speculate that simple models of evolutionary population dynamics, such as the evolution along the 1-D fitness gradient explored in \cite{Hallatschek}, are likely to exhibit fundamentally 
different behaviours from the same dynamics on neutral networks.

{\it Experimental implications.} Selection for mutational robustness at high mutation rates has been observed in both sub-viral pathogens \cite{Sanjuan,  Sole} and clonal bacterial populations \cite{Beardmore}. Recent experimental work has also shown that selection for second order effects, in this case evolvability, is observed in populations of \textit{E. coli} \cite{Lenski}. Our work suggests that, in addition to selection for robustness, populations experiencing high mutation rates in a changing environment could be subject to selection for regularity. Experiments such as those performed in \cite{Lenski}, adjusted so that the environment is periodically altered, could directly test for the selection of regularity. 
Over short periods we anticipate that successful organisms would be selected on their ability to reach their top growth rates quickly, rather than on their top growth rates themselves. 


{\it Benefit of regularity in other fields.}
For many systems in a changing environment, the ability to achieve its top performance quickly may prevail over just how good its top performance is.
For example, 
in society, a person's innate talent may be less important than the speed with which he acquires new habits or learns new skills.
In industry, companies which can quickly produce acceptable versions of desirable products may consistently outperform those which eventually produce great versions \cite{hbr}.
For living systems, where the environment is constantly changing, the ability to quickly adopt the most robust population distribution may be an essential attribute of a champion evolver.





\end{article}


\begin{thebibliography}{30}

\bibitem{Nimwegen_1999}
van Nimwegen E, Crutchfield JP, Huynen M
 (1999) {Neutral evolution of mutational robustness.}
{\it Proc Natl Acad Sci USA} 96:9716-9720.

\bibitem{Wagner_2008}
Wagner A
 (2008) {Robustness and evolvability: a paradox resolved.}
 \textit{P Roy Soc B-Biol Sci}
  275:91-100.

\bibitem{Kimura_1964}
Kimura M
 (1964) {Diffusion models in population genetics}.
 \textit{J Appl Probab} 1:177-232.

\bibitem{Kimura_1991}
Kimura M
 (1991) {Recent development of the neutral theory viewed from the
  Wrightian tradition of theoretical population genetics.}
 \textit{Proc Natl Acad Sci USA} 88:5969-5973.

\bibitem{Nimwegen_2}
van Nimwegen E, Crutchfield J P (2000)
Metastable Evolutionary Dynamics: Crossing Fitness Barriers or Escaping via Neutral Paths? 
\textit{Bull Math Bio} 62:5:799-848

\bibitem{Wright}
Wright M C, Joyce G F
 (1997) {Continuous in Vitro Evolution of Catalytic Function}
 \textit{Science} 276:614-617

\bibitem{Bershtein_2006}
Bershtein S et al.\
 (2006) {Robustness-epistasis link shapes the fitness landscape of a
  randomly drifting protein.}
 \textit{Nature} 444:929-932.
 
 
\bibitem{Maeshiro_1998}
Maeshiro T, Kimura M
 (1998) {The role of robustness and changeability on the origin and
  evolution of genetic codes}.
 \textit{Proc Natl Acad Sci USA} 95:5088-5093.

\bibitem{Tlusty}
Tlusty T
 (2008) {Casting Polymer Nets to Optimize Noisy Molecular Codes.}
 \textit{Proc Natl Acad Sci USA} 105:8238-8243

 \bibitem{Gunter_2}
Hermisson J, Wagner G P (2004) 
The population genetic theory of hidden variation and genetic robustness. \textit{Genetics} 168: 2271-2284

\bibitem{Bloom_2006}
Bloom J D, Labthavikul S T, Otey C R, Arnold F H
 (2006) {Protein stability promotes evolvability.}
 \textit{Proc Natl Acad Sci USA} 103:5869-5874.

\bibitem{Draghi_2010}
Draghi J A, Parsons T L, Wagner G P, Plotkin J B
 (2010) {Mutational robustness can facilitate adaptation.}
 \textit{Nature} 463:353-355.
 
 \bibitem{Plotkin_2}
Draghi J, Plotkin J B (2011)
Molecular evolution: Hidden diversity sparks adaptation. \textit{Nature} 474:45-46

\bibitem{Johnston_2011}
Johnston I G, Ahnert S A, Doye J P K, Louis A A
 (2011) {Evolutionary dynamics in a simple model of self-assembly}.
 \textit{Phys Rev E} 83:066105.
 
\bibitem{Ard_2}
Ahnert S E  et al.\ 
(2010) 
Self-assembly, modularity and physical complexity 
\textit{Phys Rev E} 82 026117

\bibitem{Wagner_2010}
Raman K, Wagner A
 (2010) {The evolvability of programmable hardware}.
 \textit{J Roy Soc  Interface} 8:269-281.

\bibitem{Wilke_2001}
Wilke C O et al.\
 (2001) {Evolution of digital organisms at high mutation rates leads
  to survival of the flattest.}
 \textit{Nature} 412:331-333.

\bibitem{Sole_2008}
Sardany{\'e}s J, Elena S F, Sol{\'e} R V
 (2008) {Simple quasispecies models for the survival-of-the-flattest
  effect: The role of space.}
 \textit{J Theor Biol} 250:560-568.

\bibitem{ViennaRNA_1}
Hofacker I L  et al.\
 (1994) {Fast Folding and Comparison of RNA Secondary Structures}.
 \textit{Monatsh Chem} 125:167-188.
 
\bibitem{Mandelbrot}
Mandelbrot B, {A theory of roughness lecture. 20th December 2004}.

\bibitem{Hallatschek}
Hallatschek O (2011) Noisy edge of travelling waves  \textit{Proc Natl Acad Sci USA} 108:1783  \\

\bibitem{Sanjuan} Sanju{\'a}n R et al.\ 
(2007) Selection for
robustness in mutagenized RNA viruses. {\it PLoS Genet.} 3 (6), e93, 939-946

\bibitem{Sole} Codo{\~n}er F M, Daros J A, Sole R V, Elena S F (2006) The fittest versus the flattest:
Experimental confirmation of the quasispecies effect with subviral pathogens.
{\it PLoS Pathog.} 2 (12), e136, 1187-1193

\bibitem{Beardmore}
Beardmore R E, Gudelj I,  Lipson D A,  Hurst L D
(2011) {Metabolic trade-offs and the maintenance of the fittest and the flattest}
 \textit{Nature} 472:342-346

\bibitem{Lenski}
Woods R J et al.\
(2011) {Second-order selection for evolvability in a large Escherichia coli population}
 \textit{Science} 331:1433-1436 

\bibitem{hbr}
Reeves M, Deimler M
(2011) {Adaptability: the new competitive advantage}
 \textit{Harvard Business Review} 89:3-9 

\end{thebibliography}
\end{document}